\documentstyle[aps]{revtex}
\begin{document}
\draft
\author{Nicola A. Hill}
\address{Materials Department, University of California,
Santa Barbara, CA 93106-5050}
\author{Umesh Waghmare$^{\dagger}$}
\address{Department of Physics
Harvard University,
Cambridge, MA 02138}
\title{First principles study of strain/electronic interplay in
ZnO; Stress and temperature dependence of the piezoelectric constants.}
\date{\today}
\maketitle
\begin{abstract}
We present a first-principles study of the relationship between stress,
temperature
and electronic properties in piezoelectric ZnO. Our method is a plane
wave pseudopotential implementation of density functional theory and
density functional linear response within the local density approximation.
We observe marked changes in the piezoelectric and dielectric constants
when the material is distorted. This stress dependence is the result of
strong, bond length
dependent, hybridization between the O $2p$ and Zn $3d$ electrons.
Our results indicate that fine tuning of the piezoelectric properties for
specific device applications can be achieved by control of the ZnO lattice
constant, for example by epitaxial growth on an appropriate substrate.
\end{abstract}

\input psfig

\section{Introduction}

\begin{figure}
\centerline{\psfig{figure=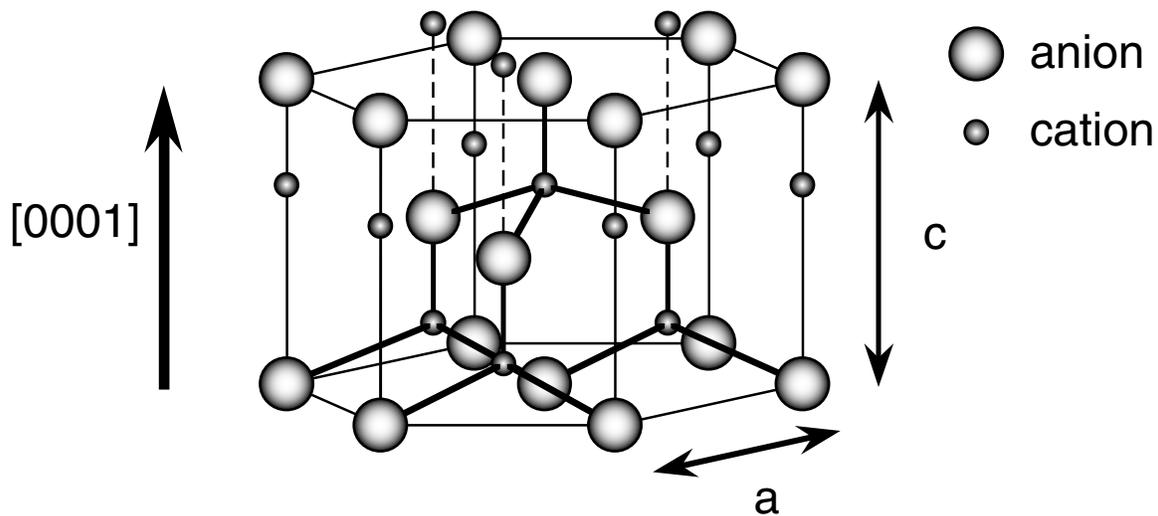,width=6.0in}}
\caption{The wurtzite structure of ZnO.}
\label{wurtzite}
\end{figure}

Zinc oxide (ZnO) is a tetrahedrally coordinated wide band gap 
semiconductor that crystallizes in the wurtzite structure
(Figure~\ref{wurtzite}).
The lack of center of symmetry, combined with a large
electromechanical coupling, result in strong piezoelectric
properties, and the consequent use of ZnO in mechanical
actuators and piezoelectric sensors. In addition, ZnO is
transparent to visible light and can be made highly conductive
by doping. This leads to applications in surface acoustic
wave devices and transparent conducting electrodes.
However the piezoelectric properties can change the 
characteristics of potential energy barriers to mobile
charges at interfaces, and hence affect the carrier
transport properties. The resulting piezoresistance
is at times desirable, for example in ZnO-based
metal-oxide varistors which can dissipate large amounts
of power in short response times and are commonly found
as electrical surge protectors\cite{Amin}. However the 
detailed effects of piezoelectrically
induced changes on the electrical behavior of ZnO 
have not yet been well characterized, and
as ZnO finds increased application in electronic devices
these effects will have large technological impact.

In this paper we present a first-principles study of the strain
dependence of the electrical properties of ZnO, using a plane
wave pseudopotential (PWPP) implementation of density
functional theory (DFT) within the local density approximation (LDA). 
We calculate and analyze the structural dependence of total
energies, band structures and piezoelectric and dielectric constants.
The principal result of our analysis is that the piezoelectric
constants of ZnO are strongly dependent on conditions of stress
and temperature, whereas the dielectric coefficients vary less
strongly.
Of particular interest is a comparison of the properties of ZnO,
in which the Zn $3d$ bands are filled
with those of PbTiO$_3$, which has empty Ti $3d$ states. Both 
materials have anomalously large
piezoelectric coefficients, but in PbTiO$_3$ the dielectric
coefficients are also anomalously large, whereas ZnO behaves
as an ordinary dielectric.
Our hypothesis is that the strong O $2p$ - Zn $3d$ hybridization
which has been previously noted in PbTiO$_3$\cite{Cohen_Krakauer,Cohen}
also occurs in ZnO, leading in both materials to strong strain-phonon
coupling and consequently large piezoelectric coefficients. Thus a filled
$3d$ band does not preclude a large peizoelectric response. The
large dielectric response in PbTiO$_3$ has a similar origin. In ZnO
however, the filled O $2p$ and Zn $3d$ bands do not allow a large
orbital response to an applied electric field, resulting in a 
normal dielectric response.

The remainder of this paper is organized as follows. In Section
~\ref{earlier_theory} we summarize the results of earlier
theoretical studies of ZnO. In Section~\ref{computational_details}
we describe the theoretical and computational methods used in this
work. In Section~\ref{bulkresults}, 
we describe our results for the static and
response properties of bulk
ZnO at its equilibrium lattice constant. 
In Section ~\ref{response} we investigate the dependence
of the electronic structure and response functions on
external stress or changes in temperature. Finally, in 
Section~\ref{Summary} we present our
conclusions and discuss implications for materials growth and
device design.

\section{Previous theoretical work}
\label{earlier_theory}

First-principles studies of ZnO are computationally challenging.
First, the wurtzite structure contains twice as many atoms per
unit cell as the zincblende semiconductor structure. In addition,
both oxygen and zinc are problematic atoms for the construction
of pseudopotentials. In both cases the relevant valence electrons
(O $2p$ and Zn $3d$) have no lower-lying electrons of the same
angular momentum to provide an effective repulsive potential
from the orthogonalization requirement\cite{Phillips}. As a
result they are tightly bound and require a large number of
plane waves in their expansion. 

The first published band structure of ZnO\cite{Rossler} used
the Green's function KKR method. This was followed by empirical
pseudopotential calculations\cite{Bloom}\cite{Chelikowsky} in which
the Zn $3d$ electrons were placed in the core. In addition to
preventing assessment of the Zn $3d$ contribution to bonding
properties, this approach was later shown to give
unsatisfactory results\cite{Schroer}. 

Schroer and co-workers\cite{Schroer} circumvented the problem
of a large plane wave basis set
by combining the use of pseudopotentials with localized Gaussian
basis sets containing orbitals of $s$, $p$, $d$ and $s^*$ symmetry.
Using this basis they compared the results of LDA calculations using
Zn$^{2+}$ (in which the $3d$ electrons are contained in the core)
and Zn$^{12+}$ (in which the $3d$ electrons are valence electrons)
pseudopotentials, and found that the Zn$^{12+}$ pseudopotential
gave results in good agreement with experiment. For example,
their LDA
energy minimum volume was 0.6 \% below the experimental value
and their calculated band structure was in reasonable agreement
with angle resolved photoemission measurements. There was a slight
discrepancy in the position of the $d$ bands which they attributed to 
inadequacy of the local density approximation in describing these
strongly correlated bands.

Dal Corso et al.\cite{dalCorso1} avoided the use of pseudopotentials 
entirely by
using the all electron full potential linear augmented plane wave (FLAPW)
method. They calculated the piezoelectric and 
polarization properties of ZnO within the LDA,
with the purpose of determining
the origin of the unusually strong piezoelectric response in ZnO.
The principal result of their work was that the contribution
to the macroscopic polarization tensor from the relative displacement
of the sublattices was large, and only partly canceled by the
electronic ``clamped-ion'' contribution, leading to a large
net piezoelectric polarization. (In contrast, in zincblende 
semiconductors these terms are of similar magnitude and opposite
sign, resulting in a small piezoelectric polarization.)
As in Ref. \cite{Schroer}, their
LDA volume slightly underestimated the
experimental value, and their Zn $d$ bands were around 4 eV higher
in energy
than observed in photoemission.

Hartree-Fock calculations have been used successfully to determine
the stability of, and transitions between, different phases of ZnO\cite{Jaffe1}
and to calculate the ($101 \bar 0$) surface reconstruction
for the wurtzite phase\cite{Jaffe2}. The Hartree-Fock approximation
also gives an incorrect
energy position for the Zn $3d$ bands - this time around 2 eV too low.

\section{Computational techniques}
\label{computational_details}

\subsection{Pseudopotential construction}

The calculations described in this work were performed using
a plane wave pseudopotential implementation\cite{Yin} of density functional
theory\cite{HKKS} within the local density approximation.
Plane wave basis sets offer many advantages in total
energy calculations for solids, including completeness,
an unbiased representation, and arbitrarily good
convergence accuracy. They also allow for straightforward
mathematical formulation and implementation, which is
invaluable in the calculation of Hellmann-Feynman forces\cite{Bendt}
and in the density functional theory linear response calculations
employed here\cite{GonzePT}.

However plane wave basis sets necessitate the use of
pseudopotentials to model the electron-ion interaction,
in order to avoid rapid oscillations of the valence
wavefunctions in the region around the ion cores. 
The difficulties 
associated with applying the pseudopotential method to 
tightly bound $d$-electrons, which
might be expected to require a prohibitively large number of plane waves to
expand their pseudopotentials, were mentioned above. 
An earlier study of cubic ZnS\cite{Martins} using the smooth
Trouiller-Martins pseudopotentials,  required plane waves up to
121 Ry in energy to achieve an energy convergence of 0.05 eV.
Although feasible for a bulk calculation for the zincblende structure
(with only two atoms per unit cell) such a large energy cutoff
is undesirable for larger unit cells, such as that of the
wurtzite structure, or those required for calculation of 
surface properties.
In this work, we 
use the optimized pseudopotentials developed by Rappe et al.\cite{Rappe},
which allow us to
reduce the required energy cutoff to 64 Ry without compromising
accuracy or transferability.
Optimized pseudopotentials minimize the kinetic energy
in the high Fourier components of the pseudo wavefunction,
leading to a corresponding reduction in the contribution of
high Fourier components in the solid. 

For both Zn and O we constructed non-relativistic optimized pseudopotentials.
The oxygen pseudopotentials were generated from a $2s^2 2p^4$ reference
configuration with core radii, $r_c$, of 1.5 a.u. for both $s$ and $p$ 
orbitals.
They were  then optimized using 4 and 3 basis functions with
cutoff wave vectors, $q_c$, of 7.0 and 6.5 a.u. for $s$ and $p$ orbitals 
respectively.
$q_c$ determines the convergence of the kinetic energy
with respect to the plane wave cutoff energy in reciprocal space
calculations. 
These oxygen pseudopotentials were used in earlier calculations
for perovskite oxides\cite{Waghmare_Rabe}\cite{HillRabe} and gave
accurate results.                            
The zinc pseudopotentials were constructed for a neutral Zn atom
with reference configuration $3d^{10} 4s^{1.75} 4p^{0.25}$.
$r_c$ values of 2.0, 1.4 and 1.4 a.u.s were used for $d$, $s$ and
$p$ orbitals respectively, with $q_c$ values of 8.0, 7.0 and 8.0 
Ry (giving a cutoff energy of 64 Ry.)
The transferability of the pseudopotential
was tested for a variety of +1 and +2 free Zn ions.
The pseudo total energies and eigenvalues were in agreement with
the all electron values to within 0.001 a.u.s.
There was no improvement in transferability, or in agreement with
all electron calculations for bulk systems, on inclusion of non linear core 
corrections\cite{NLCCs}.
All pseudopotentials were put into separable form\cite{Kleinman_Bylander}
using one projector for each angular momentum.
For both Zn and O the $l=1$ component was chosen as the local potential.
The absence of ghost states was confirmed using the ghost theorem
of Gonze, K{\" a}ckell and Scheffler\cite{Gonze}.

\subsection{Density functional theory linear response}

We use density functional theory linear response (DFT-LR) to obtain
the quadratic couplings between homogeneous strain, internal displacements
of atoms and macroscopic electric field\cite{uvw}. We use a variational
formulation of DFT-LR\cite{gat} in which the second derivative of
total energy is minimized with respect to the first derivatives of Kohn-Sham
wave functions with appropriate 
orthogonality constraints. This method avoids using
any finite-difference formulae and yields dielectric or piezoelectric
constants with a minimal number of calculations.

To obtain the piezoelectric and dielectric constants, we calculate the
DFT-LR of our system to two types of perturbations: (a) phonon (or
atomic displacements) and (b) electric field. Using the first-order
response wavefunction resulting from (a) in the Hellman-Feynman 
force formula\cite{Bendt} and in the stress formula\cite{stress},
we obtain the dynamical matrix and the 
coupling between phonons and strain respectively. 
Similarly, the response wavefunctions resulting from (b) are used to obtain
the Born effective charges and the  clamped-ion piezoelectric constants
respectively.

The symmetry of the wurtzite structure allows three independent
piezoelectric constants ($\gamma_{33}, \gamma_{13}, \gamma_{14}$) and two 
dielectric constants ($\epsilon_{33}, \epsilon_{13}$). In the present
work, we focus on $\gamma_{33}, \gamma_{13}$ and $\epsilon_{33}$. The
piezoelectric constants $\gamma_{33}$ and $\gamma_{13}$ give the 
polarization along
the $c$-axis induced by strains $e_{33}$ and $e_{11}$ respectively.
Equivalently, $\gamma_{33}, \gamma_{13}$ give the stresses 
$\sigma_{33}, \sigma_{11}$
induced on the unit cell by an electric field along the $c$-axis. The former
relationship underlies the earlier work of Dal Corso et al\cite{dalCorso1}
based on finite-difference formulae and geometric phase, 
and the latter is used in the present work based on DFT linear response.

To obtain the piezoelectric and dielectric constants,
we perform two DFT-LR calculations, one with phonon perturbations 
corresponding
to Zn and O displacements along the $c$-axis, and another with a perturbing 
electric field along the $c$-axis. The phonon perturbation allows us
to obtain  both the
frequencies of the $\Gamma$-point phonons with $z$-polarization
(which are proportional to the square roots of the
respective force constants, $\sqrt{K_{\alpha}}$), 
and the strain-phonon couplings, 
$L_{\alpha, zz}$ and $L_{\alpha, xx}$. The
LR calculations with the electric field perturbation yield the optical
dielectric constant $\epsilon_{33}^{\infty}$, the Born effective charges
$Z_{\alpha, zz}$, and the clamped-ion piezoelectric couplings
$\gamma_{33}^{0}$ and $\gamma_{13}^{0}$.
The dielectric and piezoelectric constants are then given by:

$$\epsilon_{33} = \epsilon_{33}^{\infty} + 4 \pi \sum_{\alpha} 
\frac{Z_{\alpha, zz} Z_{\alpha, zz}}{K_{\alpha}}$$
$$\gamma_{33} = \gamma_{33}^0 + \sum_{\alpha} 
\frac{Z_{\alpha, zz} L_{\alpha, zz}}{K_{\alpha}}$$
$$\gamma_{13} = \gamma_{13}^0 + \sum_{\alpha} 
\frac{Z_{\alpha, zz} L_{\alpha, xx}}{K_{\alpha}}$$.

\subsection{Technical Details}

All calculations were performed
on Silicon Graphics O2 and Origin 200 systems using
the conjugate gradient program CASTEP 2.1\cite{Payne,Castep} 
and our own related density functional linear response program\cite{Lresp}.
We used a plane wave cut
off of 64 Ry, which corresponds to around 3000 plane waves per
wavefunction  in a
a single ZnO wurtzite unit cell.
A 3x3x2 Monkhorst-Pack\cite{Monkhorst_Pack} grid was used,
leading to six k-points in the 
irreducible Brillouin Zone
for the high symmetry structures, and a correspondingly
higher number for distorted structures with lower symmetry.
The exchange correlation was parameterized using
the Perdew-Zunger parameterization\cite{Perdew_Zunger} of the
Ceperley-Alder potential\cite{Ceperley_Alder} . 

\section{Unstrained ZnO; New results and comparison with LAPW calculations}
\label{bulkresults}
                                                      
Before presenting our results for strained ZnO, we first discuss
bulk, unstrained ZnO, and compare the
results of our plane wave pseudopotential calculations with
published theoretical and experimental data. A good test of the
transferability of the pseudopotentials is that they predict
the same minimum energy structure as all-electron calculations
which use the same approximation for the exchange-correlation
functional. The minimum energy volume obtained by the earlier
all-electron LDA calculation (Ref. \cite{dalCorso1}) is 45.89 \AA$^3$.
Ref.  \cite{dalCorso1} also showed that the LDA $\frac{c}{a}$ ratio and 
$u$ value are the same as the corresponding experimental 
quantities to within 0.5 mRy, and
that the energy surface is rather flat around the minimum in the 
($u, \frac{c}{a}$) plane. 
The minimum energy volume 
obtained using our pseudopotentials is 47.6 \AA$^3$ (at the 
experimental $\frac{c}{a}$ ratio and $u$ value), and that
obtained by an earlier pseudopotential calculation using a Gaussian
basis is 46.80 \AA$^3$. The experimental volume is 47.90 \AA$^3$ and
we believe the results obtained using our method are within LDA errors. 

\subsection{Band structure analysis}

\begin{figure}
\centerline{\psfig{figure=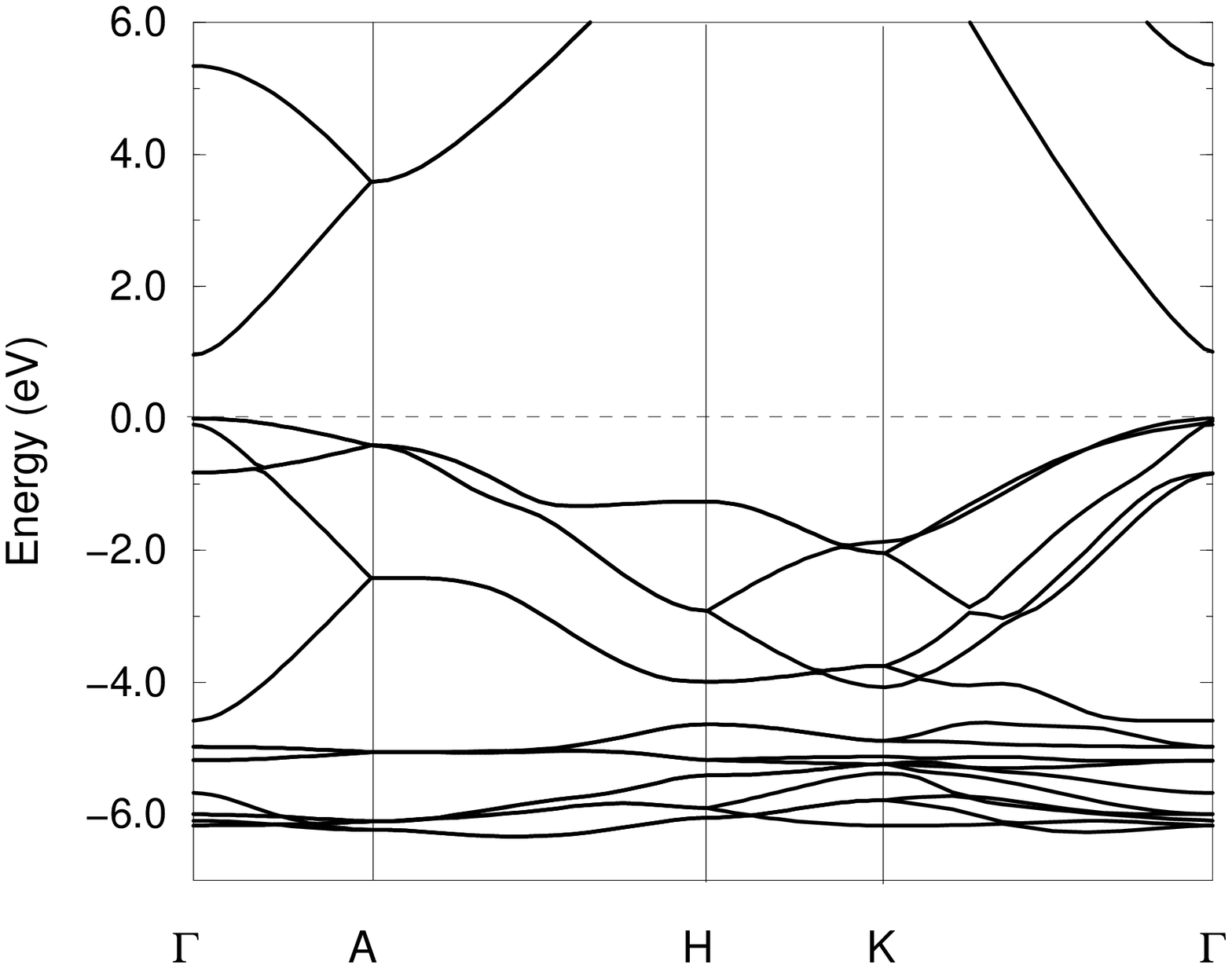,width=6.0in}}
\caption{Calculated band structure of ZnO.}
\label{ZnO_BS}
\end{figure}

In Figure~\ref{ZnO_BS} we show the calculated band structure
for ZnO at the experimental unit cell volume, ($a=3.2595$ \AA, 
$c=5.2070$ \AA\ and $u=0.3820$), which is approximately equal to
the LDA minimum volume for our pseudopotentials. 
The top of the valence band is set to 0 eV. The
band structure is indistinguishable from that presented in
Ref. \cite{dalCorso1} which was calculated using the FLAPW
technique.

\begin{figure}
\centerline{\psfig{figure=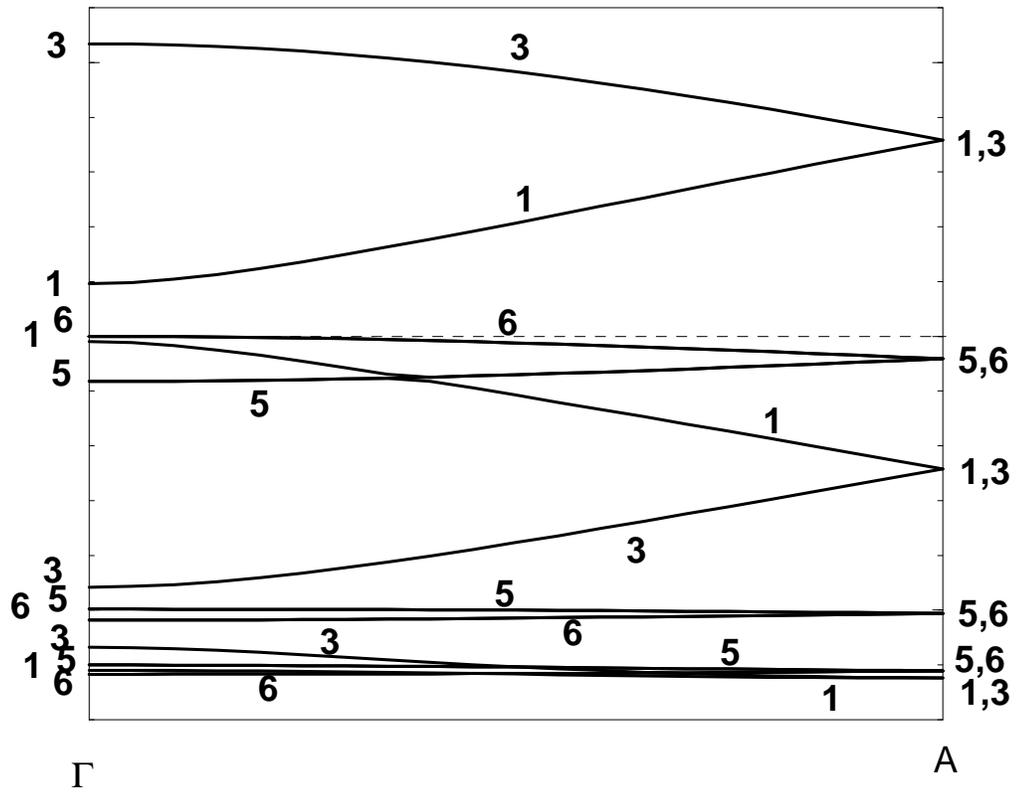,width=6.0in}}
\caption{Band structure of ZnO along the $\Gamma$ to A symmetry line
with symmetry labels added. The energy range is from -7 eV to 6eV with
the tick marks in 1 eV spacing, and the Fermi energy (0 eV) is shown by 
the dashed line. }
\label{GtoA_symm}
\end{figure}

The narrow bands between
around -6 eV and -5 eV derive largely from the Zn $3d$ orbitals
and are completely filled. The broad bands between around -5 eV and
0 eV are from the O $2p$ orbitals and are again completely filled.
The Zn $4s$ band is broad and unoccupied, ranging between 1 and 7 eV 
above the band gap.
Figure~\ref{GtoA_symm} shows the same band structure
along $\Gamma$ to A with the symmetry labels within the $C_{6v}$ group
added. 1 is the totally symmetric representation, 2 is antisymmetric
with respect to $C_6$ and $C_2$ rotation and $\sigma_v$ reflection,
5 and 6 are the doubly degenerate representations, with 5
being antisymmetric with respect to $C_6$ and $C_3$ rotation, and
6 being antisymmetric with respect to $C_3$ and $C_2$ rotation.
We see that interactions between O $2p$ and Zn $3d$ orbitals are
allowed by symmetry at $\Gamma$ and A, and along the adjoining $\Delta$ line.
The Zn $s$ orbitals can interact with O $2p$ and Zn $3d$ bands
of 1 and 3 symmetry. 

\begin{figure}
\centerline{\psfig{figure=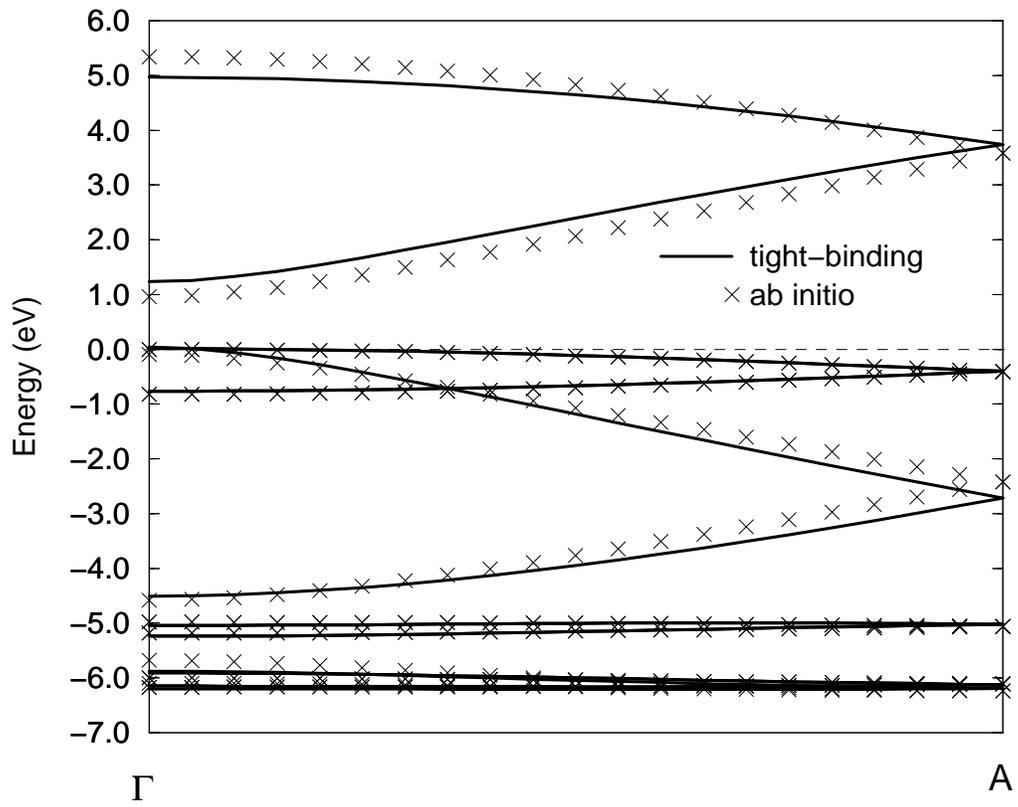,width=6.0in}}
\caption{Comparison of tight-binding and {\it ab initio}
band structures in ZnO.}
\label{GtoA_fit}
\end{figure}

In order to quantify the interactions between the various orbitals
we perform a tight-binding analysis along the $\Gamma$ to A direction
of the Brillouin zone. 
Tight-binding parameters are obtained by
non-linear-least-squares fitting\cite{Mattheiss} to the calculated
{\it ab initio} energies at the high symmetry $\Gamma$ and A points, and
at 19 points along the $\Delta$ axis.       
A good tight-binding fit (rms deviation = 0.11) is obtained when 
only nearest neighbor
interactions between O $2p$ and Zn $3d$ bands, and O $2p$ and Zn
$4s$ bands are included in the fit. Additional small Zn $3d$ -
Zn $3d$ interactions are needed to produce dispersion in the
upper $e_g$ Zn $3d$ band along this symmetry axis. The tight-binding
parameters which we obtain are given in Table~\ref{TB1} (in
the column labeled `structure 1'), and the
tight-binding band structure is compared with the {\it ab initio} 
band structure in Figure~\ref{GtoA_fit}.

The two sets of Zn-O parameters correspond to the two Zn-O distances,
the shorter ( $r_1 = 1.974$ \AA\ ) being the separation of the Zn and
O atoms lying directly above each other along the $c$ direction,
and the longer ( $r_2 = 1.989$ \AA\ )
joining Zn and O atoms in adjacent $c$-oriented
`chains'. The Zn $3d$ - Zn $3d$ interactions are small. The
Zn $4s$ - O $2p$ interactions are large, and show the expected
increase when the Zn - O spacing is decreased.
The Zn $3d$ - O $2p$ interactions are also large, and are quite
different (even changing sign) for the two different Zn - O atom pair types. 
In fact both the $\sigma$
and $\pi$ components are {\it larger} for the in-plane pairs which
have the larger Zn-O spacing. This distance dependence is unusual for
tight-binding parameters, and suggests that the nature of the Zn $3d$ -
O $2p$ hybridization is different for the Zn - O pairs lying in the
$c$ axis chains, than for the basal plane Zn - O  pairs.

\subsection{Piezoelectric and dielectric properties}

In Table~\ref{piezotab}, we present our results for piezoelectric and 
dielectric compliances and a comparison with those obtained in previous FLAPW
calculations\cite{dalCorso1} and experiments\cite{landolt}.
The agreement between the computational results is good, but the scatter in
experimental results indicates that the piezoelectric constants
may be sensitive to experimental conditions. We will investigate this
issue in the next section. We also report in Table~\ref{piezotab}
the frequencies of TO phonons with $c$ polarization
at the $\Gamma$ point, and note that only one of the TO phonons
(395 cm$^{-1}$) is IR-active; this is the only phonon which
contributes to piezoelectric and dielectric constants $\gamma_{33},
\gamma_{13}$ and $\epsilon_{33}$, studied in this work. Our results
also confirm the conclusion of Ref. \cite{dalCorso1} that the piezoelectric
constants of ZnO are dominantly contributed by phonons (internal strain).
The dielectric constant, on the other hand, has roughly equal contributions
from electrons ($\epsilon^{\infty}$) and phonons.

\section{Effects of strain and temperature on the electronic properties 
of bulk ZnO}
\label{response}

\subsection{Band structures}
\label{VA}

Next we investigate the effects of strain on the static electronic 
properties of bulk ZnO. We compare two different strained structures
with the equilibrium structure at the experimental lattice constant.
First we simulate application of a homogeneous in-plane 
compression by reducing the $a$ lattice constant by 2\%, while increasing
the $c$ lattice constant correspondingly to maintain the same 
total volume. The value of $u$ is held at 0.3820 of the $c$ axis.
Second we investigate the effect of changing the Zn-O separation
along the $c$ axis, $u$, by increasing it
by 5\% to 0.4011, while retaining the equilibrium $a$ and $c$ values.

\subsubsection{In-plane compression}

\begin{figure}
\centerline{\psfig{figure=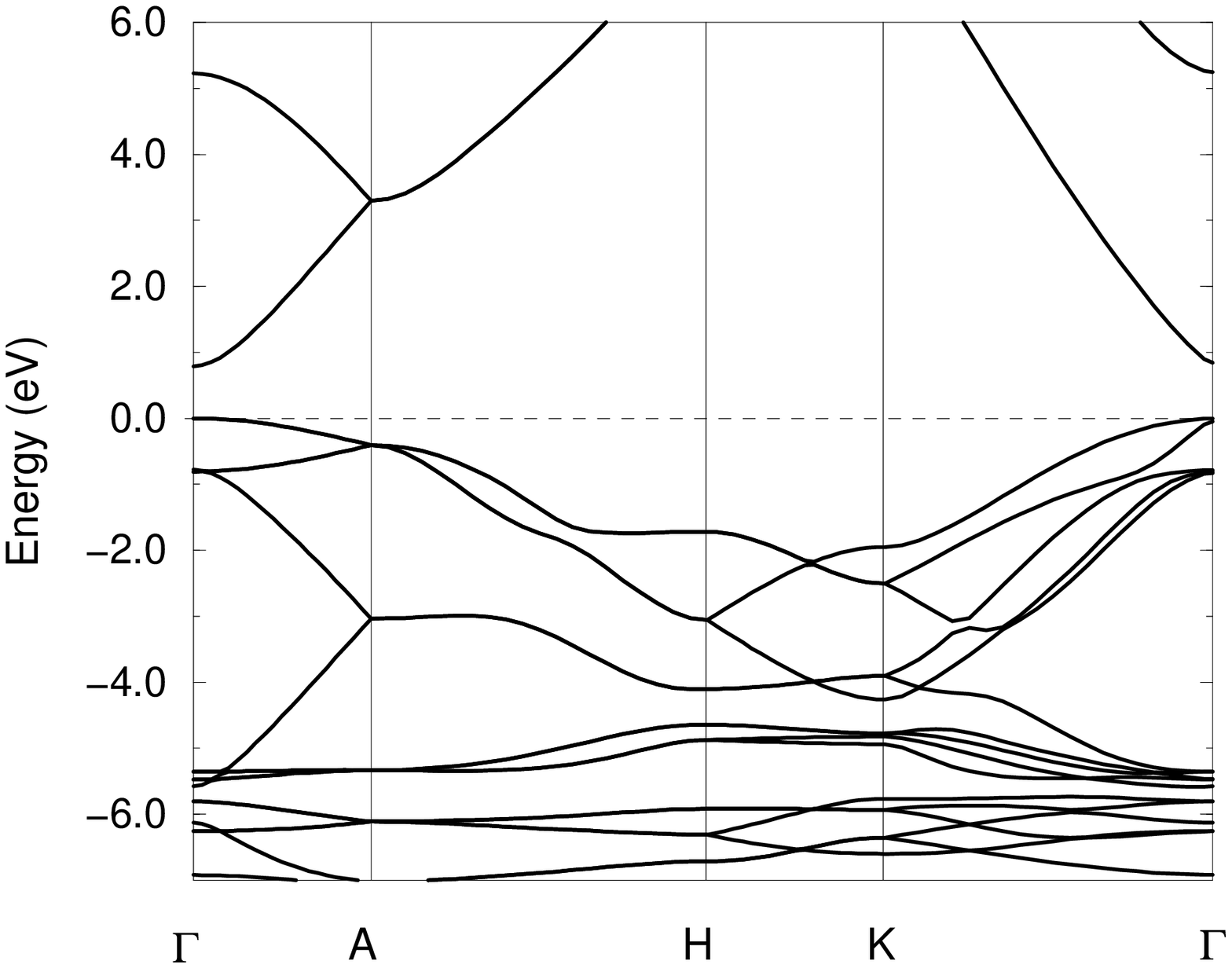,width=6.0in}}
\caption{Calculated band structure of strained ZnO.}
\label{ZnO_BS2}
\end{figure}

Application of the in-plane compression to 
reduce the $a$ lattice constant by 2\% increases the total
energy by 0.1 eV, and creates Hellman-Feynman forces on the
atoms in the $z$ direction of $\pm 0.59 \frac{\mbox{eV}}{\mbox{\AA}}$. 
The band structure of the compressed structure
is shown in Figure~\ref{ZnO_BS2}.
We observe a broadening of the bands,
as expected from the increased overlap between
the orbitals in the basal plane. 
Note in particular the broadening in
the Zn $3d$ bands. In spite of the fact that the $d$ bands are narrow,
they are very sensitive to strain and bond length as a result of 
$p-d$ hybridization processes.

In addition to an overall band broadening, there are a number of
significant details in the band structure for this compressed 
structure. First, the O $2p$ bands with symmetries 1 and 3 shift down
relative to the uppermost O $2p$ bands at the $\Gamma$ point. This
leads to a reduction in the density of states at the Fermi level,
and to an overlapping of the low energy part of these bands with 
the Zn $3d$ bands. The ordering of the lower Zn $3d$ bands is
reversed compared with those in the unstrained structure, and the
Zn $4s$ bands shift slightly down in energy, resulting in a smaller
band gap.

These observations are consistent with a tight-binding fit along
the $\Gamma$ to $A$ line, the results of which are given in Table~\ref{TB1}
(structure 2). 
Again, a good (root mean square deviation = 0.11) tight-binding fit
is obtained using the limited interaction set described above. 
Again the O $2p$ - Zn $4s$ hybridization is large and shows the
expected variation with bond length. The dependence of the (also
large) $p-d$ parameters on distance does not follow a 
straightforward pattern. Note that the anomalously low value for
the Zn $4s$ energy is the result of an attempt by the fitting
package to reproduce the down shift in the O $2p$ band of 1
symmetry (which also contains a significant Zn $4s$ component)
within our limited basis set. This energy value shifts up to a
more physical positive number if additional interactions are 
included in the basis.

\subsubsection{Change in $u$ value}

\begin{figure}
\centerline{\psfig{figure=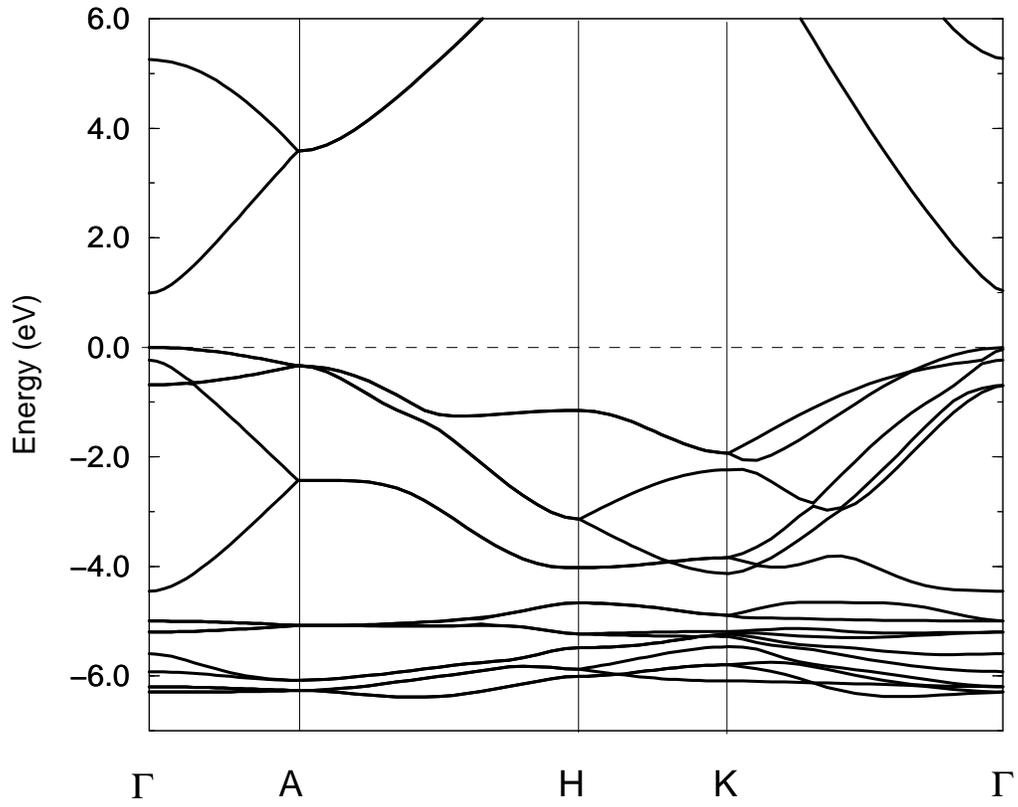,width=6.0in}}
\caption{Calculated band structure of ZnO with the $u$ value increased
by 5 \%.}
\label{ZnO_BS3}
\end{figure}

Change in the $u$ value by 5 \% increases the total energy by only
0.07 eV compared with the equilibrium structure, and introduces Hellman
Feynman forces of $\pm 0.59$ eV in the $z$ direction on the ions.
The new band structure is shown in Figure~\ref{ZnO_BS3}, and is 
qualitatively very similar to that of the undistorted structure.

A tight-binding fit using the interaction set described above produces an
rms deviation of 0.13, and the parameters listed in Table~\ref{TB1}. 
Again the O $2p$ - Zn $4s$ 
interaction is largest for the smallest bond length as expected, but
the $p-d$ hybridizations have a more complicated distance dependence.
In this case, the O $2p$ bands with 1 and 3 symmetries are shifted
down a small amount at $\Gamma$, but not as much as in the previous
structure. As a result they do not overlap with the Zn $3d$ bands. 
The Zn $4s$ bands are not shifted down relative to their position in
the unstrained case.

\subsection{Piezoelectric properties}

\subsubsection{Stress dependence}

Finally we explore the piezoelectric and dielectric properties
of strained ZnO in detail. As a result of anharmonic
couplings between different phonons and strain, the properties of
ZnO are strongly structure sensitive.
There is a large region in the phase space of structural configurations
which is low in energy and therefore contributes to the properties of ZnO.
To obtain energies, polarization
and other properties for each of these configurations from first-principles
is inefficient and impractical. 
We choose instead to use a model energy functional that captures the
physics of the low-energy structural excitations of ZnO.

To keep the model simple, we restrict our analysis to the subspace 
of degrees of freedom that preserve the symmetry of the wurtzite structure 
\{ $x=e_{xx}=e_{yy}$,
$z=e_{zz}$, $u$ \}. Justification for this simplification 
rests on the assumption (verifiable through
{\it ab initio} calculations) that most of the low
energy configurations preserve wurtzite symmetry and most of the symmetry
breaking distortions are described well at harmonic order and can be 
integrated out. This is in contrast with earlier work on 
structural phase transitions\cite{Waghmare_Rabe} where the symmetry-breaking
distortions were retained in the subspace. 

We write the model energy functional as a symmetry-invariant Taylor expansion
in atomic displacements (or normal mode degrees of freedom $u$) and strains
($x$, $z$) defined above. Including the lowest order coupling with external
stress $\sigma_{\alpha}$ and electric field along c-axis $E$,
$$E_{tot}(x,z,u, \sigma_{\alpha}, E) = \frac{1}{2} K u^2 + A u^3 + B u^4 
+\frac{1}{2} [ C_{x} x^2 + C_z z^2 + C_{xz} x z ]$$
$$ + D_x x^3 + E_x x^4
+ D_z z^3 + E_z z^4 +
 \sum_{\alpha} F_{\alpha} r_{\alpha} u^2 $$
 $$+\sum_{\alpha, \beta} G_{\alpha \beta} r_{\alpha} r_{\beta} u +
 \sum_{\alpha, \beta} H_{\alpha \beta} r_{\alpha} r_{\beta} u^2$$
\begin{equation}
- \Omega (2 \sigma_x x + \sigma_z z + 2 \gamma_{13}^0 x E + \gamma_{33}^0 z E)
-Z u E - \frac{\Omega}{4\pi}\epsilon_{\infty} E^2,
\end{equation}
where the $\gamma^0$'s and $\epsilon_{\infty}$ are clamped (or electronic)
piezoelectric and dielectric constants, and 
$r_{\alpha}$ is strain $x$ or $z$. $K$, $A$ ... 
$H$, $\gamma$ and $\epsilon$ are the harmonic and anharmonic
coupling parameters including force, elastic and mixed coupling
constants. These parameters have been determined from DFT total energy and linear
response calculations.

The equilibrium state of ZnO under the application of external stress or
electric field at zero kelvin is obtained by minimizing the total energy 
$E_{tot}$ with 
respect to the structural parameters $x$, $z$ and $u$:
\begin{equation}
\frac{ \partial E_{tot}}{\partial x} = 0, \frac{ \partial E_{tot}}{\partial z}=0,
\frac{ \partial E_{tot}}{\partial u}=0.
\end{equation}
For the equilibrium structure, the static
dielectric and piezoelectric constants are calculated using 
the expressions in Section~\ref{computational_details}  
(also in Ref. \cite{uvw}), 
with various coupling parameters being dependent on the structure.
The spontaneous polarization is calculated using the expression
$$P = -\frac{ \partial E_{tot}}{\partial E}.$$

\begin{figure}
\centerline{\psfig{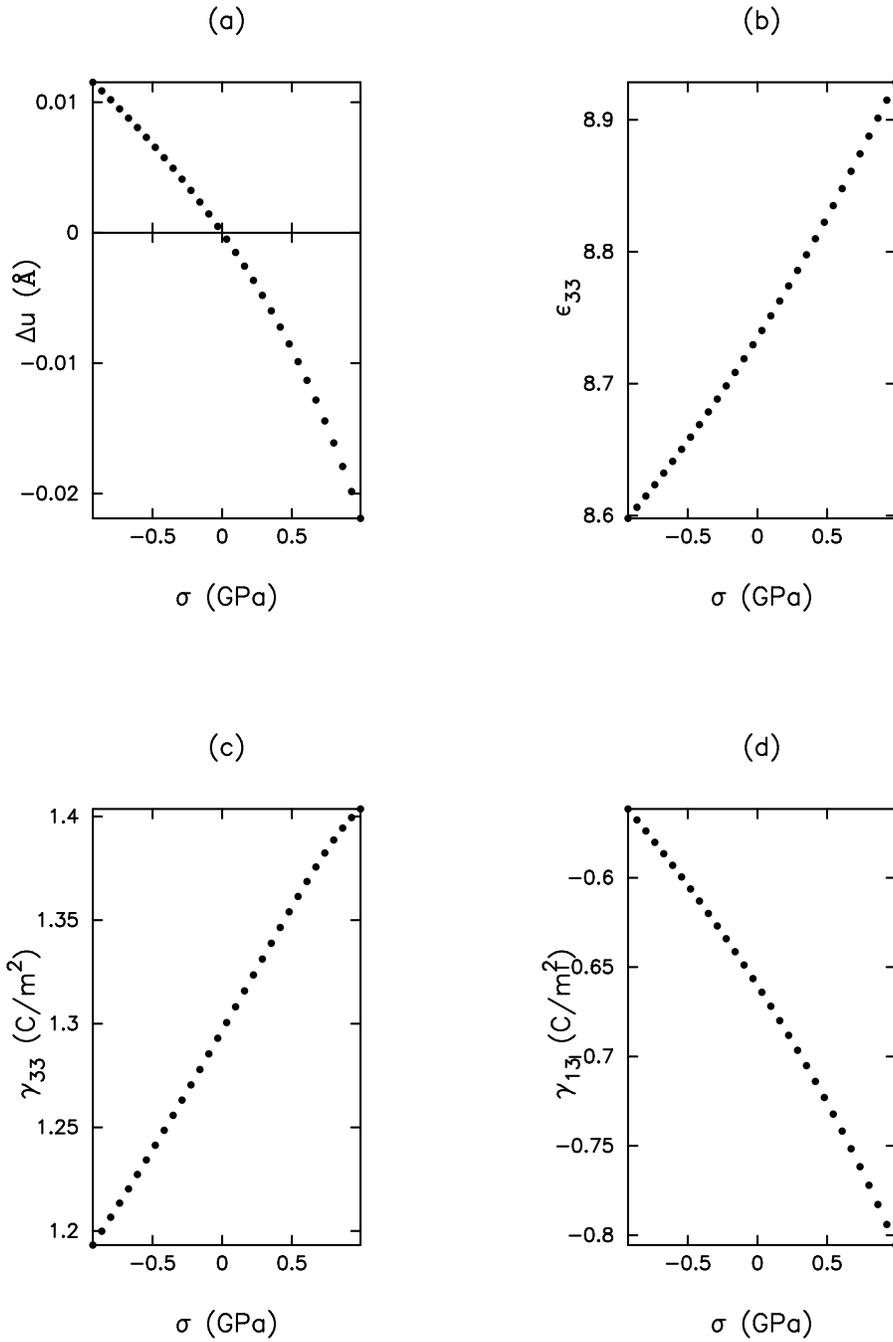}}
\caption{
Dependence of the change in structural parameter $\Delta u$ (a),
dielectric $\epsilon_{33}$ (b), piezoelectric constants $\gamma_{33}$
(c) and $\gamma_{13}$ (c) of ZnO on the applied stress in the basal
plane $\sigma=\sigma_{xx}=\sigma_{yy}$.
}  
\label{stress_dep}
\end{figure}

ZnO is grown in the form of thin solid films on sapphire
and there are always interfacial stresses in the films. We
use our model to study properties of ZnO as a function of 
stress in the basal $ab$ plane. In Fig. \ref{stress_dep}, we
show our results for the structural parameter $u$, dielectric
and piezoelectric response of  ZnO as the applied
stress $\sigma=\sigma_{xx}=\sigma_{yy}$ is varied from -1 to
1 GPa. 

We find that parameter $u$ changes by about 1 percent in this
range of basal stress. The dielectric constant $\epsilon_{33}$ 
monotonically increases by about 2 to 3 percent. In contrast,
the piezoelectric constants are rather sensitive to stress
changing by about 15 to 30 \%. With access to different contributions
to these constants in our model, we discover that most of
the dependence of these compliances on stress is due to the
phonon contribution. 

\subsubsection{Temperature Dependence}

ZnO is also a pyroelectric material, characterized by a temperature
dependence of the polarization. This has technological relevance
because it leads to the widespread use of ZnO in
infra-red detectors.
To explore various stress and electric field-dependent properties of ZnO
at finite temperature, we obtain free energy functional using a local harmonic
model\cite{lesar} for entropy. In this model, entropy is calculated treating 
phonons harmonically for given structural parameters. In the present work, we 
include the optical phonons that are polarized along $z$-axis. With these
approximations, the free energy is:
\begin{equation}
G(x,z,u, \sigma_x, \sigma_z, E, T) = E_{tot}(x,z,u, \sigma_x, \sigma_z, E)
+3 k_B T Log( \frac{ (K_1 K_2 K_3)^{\frac{1}{3}} } {k_B T} )
\end{equation}
where the $K_i$'s are the harmonic force
constants of the three $\Gamma-$phonons with $z$-polarization for given values
of structural parameters. Due to the anharmonic terms included in the energy
expansion Eqn (1), $K_i$'s (hence the T-dependent part of the free energy)
depend on the structure.
The equilibrium state of ZnO under the application of external stress or
electric field at finite temperature
is then obtained by minimizing the free energy $G$ with 
respect to the structural parameters $x$, $z$ and $u$:
\begin{equation}
\frac{ \partial G}{\partial x} = 0, \frac{ \partial G}{\partial z}=0,
\frac{ \partial G}{\partial u}=0.
\end{equation}
Again, the 
dielectric and piezoelectric constants are calculated using 
the expressions in Section ~\ref{computational_details}, and
the spontaneous polarization is now calculated using the expression
$$P = -\frac{ \partial G}{\partial E}.$$

\begin{figure}
\centerline{\psfig{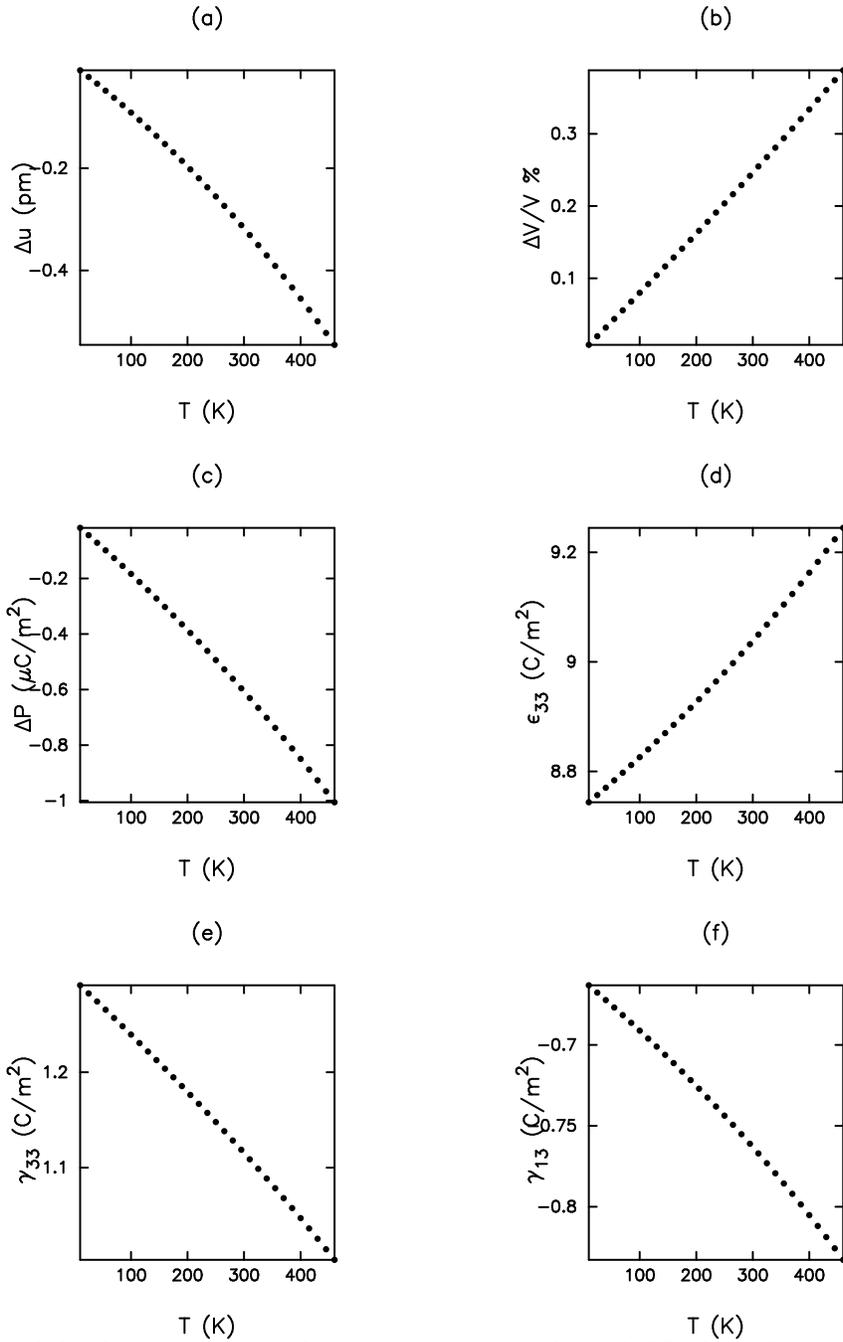}}
\caption{
Dependence of the change in structural parameter $\Delta u$ (a), volume
$\Delta V/V$ (b) and spontaneous polarization $\Delta P$ (c),
dielectric constant $\epsilon_{33}$ (d), piezoelectric constants
$\gamma_{33}$ (e) and $\gamma_{13}$ (f) of ZnO on the
temperature.
}
\label{temp_dep}     
\end{figure}

We investigate the dependence of various properties of ZnO
on temperature. In Fig. \ref{temp_dep}, we display results for
structural parameters, dielectric and piezoelectric constants,
and polarization for a range of temperatures from 0 to 450 K.
We find that the structural parameters such as the bond-length $u$
and volume change only by about 0.3 \% from zero kelvin to room temperature.
 
The pyroelectric constant, which 
we obtain from our calculated results for polarization, is 20 $\mu C/m^2/K$, 
compared with the experimental value of 9.4 $\mu C/m^2/K$\cite{landolt}.
Considering the simplicity in our treatment of temperature through
the model entropy function, we find this agreement quite encouraging.
In particular, we point out that, while the nonpolar TO phonons 
with $z$-polarization
have been omitted from the expansion of energy,
they {\it have} been included in the entropy term, where we use
the expression of entropy\cite{lesar} treating phonons harmonically,
consistent with our energy expansion.

To estimate the effect of these non-polar phonons on the pyroelectric constant,
we omitted their contribution to entropy and found 
a pyroelectric constant of about 40 $\mu C/m^2/K$ (almost doubled).
If we assume that phonons should generally suppress the pyroelectric
constant, the discrepancy between theory and experiment is likely due to our
omission of phonons with $x$ and $y$-polarization from the entropy expression.

The linear dielectric response $\epsilon_{33}$
of ZnO is quite sensitive to temperature, changing
by about 4\% in the temperature
range considered. The piezoelectric response, on the other hand, is very
sensitive to temperature changing by about 20\%. This should be an
important consideration in designing piezoelectric devices for
operation at room temperature.

\subsection{Discussion}

\begin{figure}
\centerline{\psfig{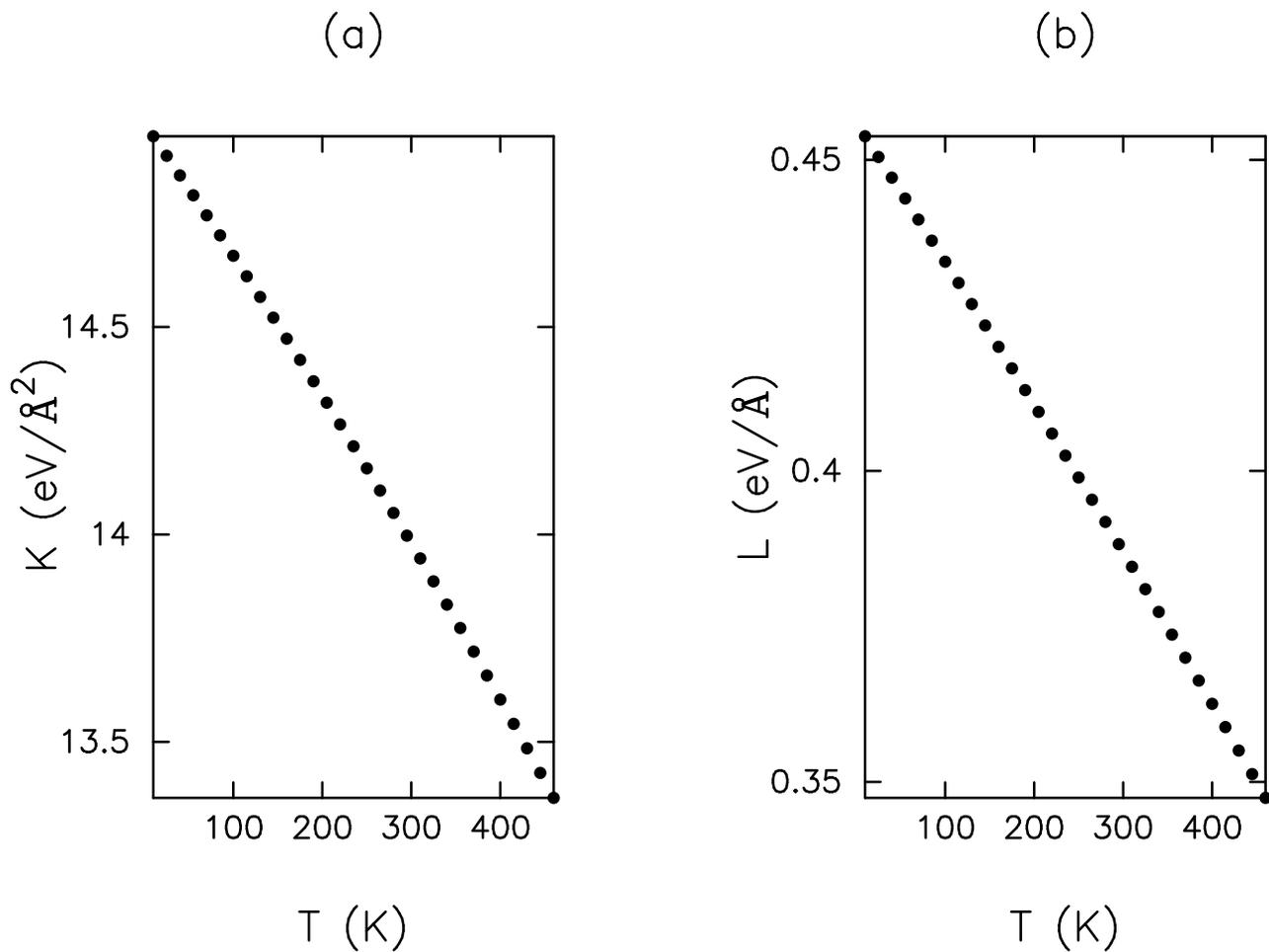}}
\caption{
Dependence of (a) the force constant and (b)coupling with strain 
of the polar TO phonon as a function of temperature.
}
\label{coup_dep}
\end{figure}

It is clear from our results that the piezoelectric response of
ZnO is strongly sensitive to both temperature and stress, changing 
by up to 30 \% over the range of parameters considered. This dependence 
arises from the changes in structural parameters (manifested
through the phonon contribution). We saw in 
Section~\ref{computational_details} that the
phonon contribution to the 
piezoelectric constants arises from the coupling of phonons with strain, $L$,
and Born effective charge, $Z$,
$$\gamma_{phonon} =   \frac{L\cdot Z}{K},$$
$K$ being the force constant.
In Fig. \ref{coup_dep}, we show how $K$ and $L$ for the polar TO phonon
change with temperature. While $K$ changes by only 10 \%, the coupling with
strain, $L$, changes by about 25 \%. The Born effective charge $Z$ 
(which describes the coupling
of this phonon with electric field) is not found to vary much with
temperature. The large temperature dependence of the piezoelectric response 
arises predominantly from that of the coupling of phonon with strain and
its force constant. Since only the latter contributes to the
dielectric constants, dielectric properties are less sensitive to 
structure, stress or temperature.

In the Section~\ref{VA}, we found the hybridization between Zn $d$ orbitals
and O $p$ orbitals to be sensitive to structural parameters. The same 
hybridization was
found to be the cause of anomalous Born effective charges in ferroelectric
materials such as PbTiO$_3$\cite{Cohen}. 
While the $d$ orbitals of the transition metals in perovskite
ferroelectrics are formally unoccupied, those in Zn are fully occupied, leading
to normal effective charges. The coupling of phonons with strain, however, 
is large and structure dependent irrespective of the
occupancy of $d$ orbitals.

\section{Summary}
\label{Summary}

In summary, we have calculated the
electronic and atomic structure of ZnO from first-principles,
and analyzed the nature of the bonding using the
tight-binding method. We find that
hybridization between the Zn $d$ orbitals and O $p$ orbitals is
strongly structure
dependent. Using DFT linear response, we have obtained the phonon frequencies,
dielectric and piezoelectric constants of ZnO at zero temperature, and
have shown that phonons (internal strain)
 have the dominant contribution to piezoelectricity in ZnO.
>From DFT linear response and total energy calculations, and a simple
model for vibrational entropy, we have 
constructed an {\it ab initio} free energy
functional for ZnO to study its properties at finite temperature and
under applied stress. Our results show that the piezoelectric properties
of ZnO are strongly dependent on both temperature and stress. 
This clearly has implications for the design of devices intended to
operate at room temperature, or under stressed conditions.
By analyzing various
physical contributions, we have found that this is primarily due 
to the coupling
between phonons and strain. The O $2p$ - Zn $3d$ hybridization
is the cause of the large magnitude and sensitivity
of this coupling.

\section{Acknowledgments}

The authors thank Andrew Rappe and Nicholas Ramer for their assistance
in generating the Zn pseudopotential, and Paul Verghese and V. Srikant 
for many useful discussions. Many of the ideas presented in this paper 
were formulated during the 1998 ``Physics of Insulators'' workshop at the 
Aspen Center for Physics.
N.A. Hill's funding was provided through
the NSF-POWRE program, grant \# DMR-9973859. U. V. Waghmare's 
funding was provided by the Office of Naval Research, grant 
\# N00014-97-1-1068.

\clearpage

%
%

\begin{table}
\begin{center}
\begin{tabular}{|| l | r  | r  | r ||} 
property                 &  Ref. \cite{dalCorso1}  &  Present Work &  Experiment \cite{landolt} \\ \hline
$Z^{\star}_{Zn, z}$      &     2.05                &  2.07         &  2.10 \\
$\gamma_{33}^0$ $(C/m^2)$&    -0.58                & -0.73         &   -  \\
$\gamma_{33}$  $(C/m^2)$ &     1.21                &  1.30         &  1.0-1.5 \\
$\gamma_{13}^0$ $(C/m^2)$&     0.37                &  0.31 &    -    \\
$\gamma_{13}$  $(C/m^2)$ &    -0.51                & -0.66 &    -0.36 to -0.62\\
$\epsilon_{33}^{\infty}$ &       -                &  4.39  &   -  \\
$\epsilon_{33}$          &       -                &  8.75  &   -  \\
TO phonon $(cm^{-1})$    &       -                &  544.919, 395.349, 258.519 &  -  

\end{tabular}
\end{center}
\caption{Comparison between results of FLAPW calculations 
(Ref. \protect \cite{dalCorso1}), pseudopotential
calculations (this work)
and experimental results (Ref.\protect \cite{landolt})
for piezoelectric constants and related 
properties of ZnO.}
\label{piezotab}
\end{table}

\begin{table}
\begin{center}
\begin{tabular}{|| l | r | r | r ||} 
parameter                & structure 1 & structure 2 & structure 3 \\ \hline
$E_{O2p}$                & -1.262 & -1.062  & -1.308 \\
$E_{Zn4s}$               &  1.235 & -0.274  &  1.409 \\
$E_{Zn3d}$               & -5.144 & -5.396  & -5.159 \\
$V_{O2p-Zn4s}^1$         &  2.584 &  2.936  &  2.823 \\
$V_{O2p-Zn4s}^2$         &  2.386 &  2.517  &  2.279 \\
$V_{(O2p-Zn3d)\sigma}^1$ &  0.625 &  0.829  &  0.591 \\
$V_{(O2p-Zn3d)\sigma}^2$ & -0.856 & -1.343  & -0.761 \\
$V_{(O2p-Zn3d)\pi}^1$    &  1.292 &  1.444  &  1.294 \\
$V_{(O2p-Zn3d)\pi}^2$    & -1.711 & -1.365  & -1.802 \\
$V_{(Zn3d-Zn3d)\sigma}$  &  0.169 &  0.124  &  0.170 \\
$V_{(Zn3d-Zn3d)\pi}$     & -0.006 &  0.001  & -0.014 \\
$V_{(Zn3d-Zn3d)\delta}$  &  0.053 &  0.034  &  0.060
\end{tabular}
\end{center}
\caption{Tight-binding parameters (in eV) for ZnO
obtained by non-linear-least-squares fitting to the {\it ab initio}
eigenvalues along $\Gamma$ to A. E indicates an orbital energy,
and V an inter-atomic transfer integral. The transfer integrals
with the superscript `1' are between the closest nearest neighbor
Zn-O pairs, and those with the superscript `2'
are between the nearest neighbors with the larger separation. 
Only the parameters listed
in the table were allowed to be non-zero in the fitting procedure.}          
\label{TB1}
\end{table}

\begin{references}

\bibitem[ \dagger ] {} Present Address: Theoretical Sciences Unit,
J. Nehru Centre for Advanced Scientific Research, Jakkur, Bangalore,
560 064, India.

\bibitem{Amin}
A. Ahmin, J. Am. Ceram. Soc. {\bf 72}, 369 (1989).

\bibitem{Cohen_Krakauer}
R.E. Cohen and H. Krakauer, Ferroelectrics {\bf 136}, 95 (1992).

\bibitem{Cohen}
R.E. Cohen, Nature {\bf 358}, 136 (1992).
                                                  
\bibitem{Phillips}
J.C. Phillips and L. Kleinman, Phys. Rev. {\bf 116}, 287 (1959).

\bibitem{Rossler}
U. Rossler, Phys. Rev {\bf 184}, 733 (1969).

\bibitem{Bloom}
S. Bloom and I. Ortenburger, Phys. Stat. Sol B {\bf 58}, 561 (1973).

\bibitem{Chelikowsky}
J.R. Chelikowsky, Sol. Stat. Comm. {\bf 22}, 351 (1977).

\bibitem{Schroer}
P. Schr\"{o}er, P. Kr\"{u}ger and J. Pollmann, Phys. Rev. B {\bf 47},
6971 (1993).

\bibitem{dalCorso1}
A. Dal Corso, M. Posternak, R. Resta and A. Baldereschi,
Phys. Rev. B {\bf 50}, 10715 (1994).

\bibitem{Jaffe1}
J.E. Jaffe and A.C. Hess, Phys. Rev. B {\bf 48}, 7903 (1993).

\bibitem{Jaffe2}
J.E. Jaffe, N.M. Harrison and A.C. Hess, Phys. Rev. B {\bf 49}, 
11153 (1994).

\bibitem{Yin}
M.T. Yin and M.L. Cohen, Phys. Rev. B {\bf 26}, 5668 (1982).

\bibitem{HKKS}
H. Hohenberg and W. Kohn, Phys. Rev. {\bf 136}, 864 (1964);\\
W. Kohn and L.J. Sham, Phys. Rev. {\bf 140}, 1133 (1965).
                                                                            
\bibitem{Bendt}
P. Bendt and A. Zunger, Phys. Rev. Lett. {\bf 50}, 1684 (1983).

\bibitem{GonzePT}
X. Gonze. Phys. Rev. B {\bf 55}, 10337 (1997).

\bibitem{Martins}
J.L. Martins, N. Troullier, and. S.-H. Wei, Phys. Rev. B {\bf 43},
2213 (1991).

\bibitem{Rappe}
A.M. Rappe, K.M. Rabe, E. Kaxiras and J.D. Joannopolous,
Phys. Rev. B {\bf 41}, 1227 (1990).

\bibitem{Waghmare_Rabe}
U.V. Waghmare and K.M. Rabe, Phys. Rev. B {\bf 55}, 6161 (1997).

\bibitem{HillRabe}
N.A. Hill and K.M. Rabe, Phys. Rev. B {\bf 59}, 8759 (1999).

\bibitem{NLCCs}
S.G. Louie, S. Froyen and M.L. Cohen, Phys. Rev. B {\bf 26}, 1738 (1982).

\bibitem{Kleinman_Bylander}
L. Kleinman and D.M. Bylander, Phys. Rev. Lett. {\bf 48}, 1425 (1982).

\bibitem{Gonze}
X. Gonze, P. K{\" a}ckell and M. Scheffler, Phys. Rev. B {\bf 41},
12264 (1990).

\bibitem{uvw} U. V. Waghmare, unpublished.

\bibitem{gat} X. Gonze, D. C. Allan and M. P. Teter, Phys. Rev. Lett.
{\bf 68}, 3603 (1992).

\bibitem{stress} O. H. Nielsen and R. M. Martin, Phys. Rev. {\bf B32},
3792 (1985).

\bibitem{Payne}
M.C. Payne, M.P. Teter, D.C. Allan, T.A. Arias and J.D. Joannopoulos,
Rev. Mod. Phys. {\bf 64}, 1045 (1992).

\bibitem{Castep}
M.C. Payne, X. Weng, B.Hammer, G. Francis, U. Bertram, A. de Vita,
J.S. Lin, V. Milman and A. Qteish, unpublished.

\bibitem{Lresp}
U. V. Waghmare, K. M. Rabe and V. Milman, unpublished.

\bibitem{Monkhorst_Pack}
H.J. Monkhorst and J.D. Pack, Phys. Rev. B {\bf 13}, 5188 (1976).

\bibitem{Perdew_Zunger}
J.P. Perdew and A.Zunger, Phys. Rev. B {\bf 23}, 5048 (1981).

\bibitem{Ceperley_Alder}
D.M. Ceperley and B.J. Alder, Phys. Rev. Lett. {\bf 45}, 566 (1980).

\bibitem{Mattheiss}
L.F. Mattheiss, Phys. Rev. B {\bf 6}, 4718 (1972).

\bibitem{landolt} {\it Landolt-B\"ornstein:
Numerical data and functional relationships in science and technology,
} Vol. III/11, ed. by K. H. Hellwege and A. M. Hellwege, (Springer,
1979).

\bibitem{lesar} R. LeSar, R. Najafabadi and D. J. Srolovitz,
Phys. Rev. Lett. {\bf 63}, 624 (1989).

\end{references}
\end{document}